\documentclass[%
 aip,
 amsmath,amssymb,
 reprint,%
]{revtex4-1}

\usepackage{graphicx}
\usepackage{dcolumn}
\usepackage{bm}
\usepackage{float}
\usepackage{xcolor}
\usepackage[utf8]{inputenc}
\usepackage[T1]{fontenc}
\usepackage{mathptmx}
\usepackage{etoolbox}
\usepackage{placeins}

\newcommand{\new}[1]{\textcolor{black}{#1}}

\makeatletter
\def\@email#1#2{%
 \endgroup
 \patchcmd{\titleblock@produce}
 {\frontmatter@RRAPformat}
 {\frontmatter@RRAPformat{\produce@RRAP{*#1\href{mailto:#2}{#2}}}\frontmatter@RRAPformat}
 {}{}
}%
\makeatother

\begin{document}


\title[The effects of resistivity on oscillatory reconnection and consequences for solar flare Quasi Periodic Pulsations]{The effects of resistivity on oscillatory reconnection and consequences for solar flare Quasi Periodic Pulsations} 



\author{Luiz A. C. A. Schiavo}
\email[]{luiz.schiavo@northumbria.ac.uk}
\affiliation{Department of Mathematics Physics \& Electrical Engineering, Northumbria University, Ellison PI, Newcastle upon Tyne, NE1 8ST, UK}
\affiliation{Department of Physics and Astronomy, University of Manchester, Oxford Road, Manchester, M13 9PL, UK}
\author{James Stewart}
\email[]{james.stewart@manchester.ac.uk}
\affiliation{Department of Physics and Astronomy, University of Manchester, Oxford Road, Manchester, M13 9PL, UK}

\author{Philippa K. Browning}
\email[]{philippa.browning@manchester.ac.uk}
\affiliation{Department of Physics and Astronomy, University of Manchester, Oxford Road, Manchester, M13 9PL, UK}


\date{\today}

\begin{abstract}
Quasi-periodic pulsations (QPPs) are often observed in flare emissions. While these may reveal much about the time-dependent reconnection involved in flare energy release, the underlying mechanisms are still poorly understood. In this paper, we use 2D magnetohydrodynamic simulations to investigate the magnetic reconnection in two merging flux ropes, focusing on the effects of the resistivity on the time variation of the reconnection. We consider both uniform resistivity and current-dependent anomalous resistivity profiles. Our findings reveal that resistivity plays a critical role in controlling the reconnection dynamics, including \new{reconnection rate} oscillations and the rate of decay of the reconnection rate. Resistivity \new{also influences} the oscillations in emitted gyrosynchrotron radiation. However, in contrast to this strong influence of resistivity on reconnection rates, we observed a different behaviour for the emitted waves, whose frequencies are almost independent of resistivity variations.
\end{abstract}

\pacs{}

\maketitle 

\section{Introduction}
Solar flares are dramatic releases of stored magnetic energy in the solar corona, through the process of magnetic reconnection. They are observed as strong brightenings in emission across the electromagnetic spectrum, including soft and hard X-rays and radio from coronal hot plasma and non-thermal electrons and ions, on timescales of minute to hours. Flares are of considerable physical interest as significant phenomena in the solar atmosphere, but they also have significant practical consequences as being the drivers of space weather. Furthermore, flares are widely observed on other stars\citep{Benz2010}, and a better understanding of their solar counterparts will have impact on understanding these and on the implications for any planets orbiting these stars.  Current observational and theoretical understanding of flares is summarised in review papers \cite{Benz2008,Fletcher2011,Shibata2011,Janvier2015}. 

Flares are complex phenomena, and although it is now generally accepted that magnetic reconnection is the ultimate cause of the energy release, many questions about their dynamics remain unanswered. Reconnection itself is a fundamental physical process occurring across the universe, and is also a subject of intensive research, with current understanding now moving far beyond the classical models of steady, two-dimensional reconnection in a resistive fluid \citep{[e.g]Priest2000,Ji2022,Yamada2022}. Improved modelling of reconnection is vital in order to better understand flares. Here, the focus is on time-dependent reconnection - in contrast with classical steady  reconnection models such as Sweet-Parker and Petschek.

One feature of observed solar and stellar flare emission which is not well understood is the presence of Quasi-period pulsations (QPPs). These are oscillations widely observed in the Extreme Ultraviolet, soft and hard X-ray and radio light curves, with periods around 1 - 100 s and amplitudes around 1-10$\%$. While the periods suggest that the QPPs are fundamentally a magnetohydrodynamic (MHD) phenomenon, there is no consensus around the mechanism(s) producing them, and a number of processes are currently proposed, as discussed in recent reviews \cite{McLaughlin2018,Kupriyanova2020-lp,Zimovets2021}. A better understanding of the origins and nature of QPPs could shed light on the nature of reconnection and energy release in flares, and allow the pulsations to be used as a seismological diagnostic of the physical conditions in the flaring plasma \cite{Karampelas2023}. A key question is whether the timescales of the pulsations arises due to the time-dependent nature of the energy release process (for example, through oscillatory reconnection), or is an intrinsic oscillation of the background magnetic field (as in the sound from a plucked guitar string), or is determined by some external wave propagating into the flaring region. The main aim of this paper is to distinguish between these possibilities, using magnetohydrodynamic simulations of reconnection.

Oscillatory magnetic reconnection is a strong candidate for causing pulsations in flare emission, and is likely to play some role in flares since they are clearly transient events and any reconnection must be time-dependent. Oscillations in the reconnection process may arise intrinsically or through modulation from some external driver. Oscillatory reconnection was shown by \citet{McLaughlin2004,Karampelas2022,Karampelas2023} to arise when an incoming fast wave pulse impinges on a magnetic X-point. We also consider oscillatory magnetic reconnection at an X-point, but our work differs in some significant respects: notably,  we include a guide field which is more representative of solar flares magnetic field configurations, and we do not impose  an  external driving pulse but rather the reconnection arises naturally from the initial magnetic field.  Merging twisted magnetic flux ropes can lead to oscillatory reconnection \cite{Knoll2006,Stanier2013,Stewart2022}, as well as the slow outward propagation of wave-like disturbances from the reconnection site after the merger has been completed \cite{Stewart2022}. It has been  proposed that this oscillatory behaviour, also seen in a 0D analogue model for flux rope coalescence \cite{Kolotkov2016},  could provide a mechanism for generating QPPs in flares, a hypothesis which we investigate further here. 

The merging flux rope scenario provides a means to study reconnection in a configuration with free  energy associated with currents, as must exist in a solar flare \cite{Tajima1987,Sakai}. It is also of much wider relevance within plasma physics, occurring, for example, during some formation schemes for spherical tokamaks \cite{Stanier2013,Gryaznevich2017,Tanabe2017,Ahmadi2021}. Many studies of generic  reconnection physics in the  merger of magnetic islands (which are flux ropes in the presence of a guide field) have been performed using the Particle-in-Cell, hybrid, Hall-MHD and other approaches \cite{Stanier2015,Stanier2017,Du2018,Makwana2018} as well as 2D and 3D MHD \cite{Biskamp1980,Tam2015,Huang2016,Beg2022}. The merger of two (or more flux ropes) into a single flux rope may be viewed as a relaxation process towards a minimum energy state with an inverse cascade from smaller to large length scales \cite{Browning2014,Browning2016,Robinson2023}.

One way in which flux rope merger may naturally arise is through coalescence instability. This occurs in a chain of magnetic islands, whereby a small disturbance leads to a force imbalance in which neighbouring islands are displaced towards each other and mutually attract \cite{Finn1977,Biskamp1980}. In the nonlinear phase of the instability, islands will merge through magnetic reconnection. If there is a guide field (an out-of-plane magnetic field component), the islands are actually twisted magnetic flux ropes. This phenomenon has become of considerable interest more recently due to the discovery of the plasmoid instability, in which a long reconnection current sheet fragments into plasmoids (magnetic islands or  - in the presence of a guide field - flux ropes)\cite{Shibata2001,Loureiro2007}. Whilst originally postulated within a 2D framework, it is now evident that plasmoids form within 3D magnetic fields as well, and a complex configuration of interacting 3D flux ropes may develop\cite{Archontis2006,Mulay2023}. There is increasing observational evidence for the presence of plasmoids within coronal current sheets associated with solar flares \cite{Milligan2010,Takasao2012,Kumar2013,Dai2018,Yan2022,Kumar2023} as well as near the terrestrial magnetopause \cite{Wang2020}.

In the solar corona, flux rope merger on more global scales may play a key role in solar flares and other eruptive events, involving the interactions of large-scale twisted loops \cite{Tajima1987,Sakai}.  Recently, flux rope merger has been proposed as a cause for the magnetic switchbacks observed in the solar wind by Parker Solar Probe \cite{Agapitov2022}. The MHD simulations presented here may be appropriately scaled to represent either the plasmoid coalescence within a flaring current sheet or merger of large scale twisted coronal loops or flux ropes. 

We build on the model of \citet{Stewart2022} in order to answer a crucial question about the origin of QPPs. Are the periodicities of the observed oscillations determined by the time-variations of the underlying energy release process (magnetic reconnection) or are they determined by the natural frequencies of MHD modes in the ambient magnetic configuration?  As reconnection is naturally affected by resistivity, whereas natural oscillations are not, studying the dependence on resistivity provides a key to answering this question. We also investigate the impact of the resistivity profile on the dynamics of our system and the resulting QPP-like oscillations. \citet{Talbot_2024}, considering a different oscillatory reconnection scenario,  indicated that resistivity can influence the maximum amplitude of the current density in oscillatory reconnection, the nature of the decay rate, and the magnitude of Ohmic heating at the null point, potentially altering the dynamics of the  system. In order to fully understand QPPs, it is necessary  to link models  with observations, and we therefore, for the first time, use a forward-modelling approach to predict the observable microwave emissions arising from our model. 

Our work conclusively demonstrates that, although oscillatory reconnection is present, it acts simply as a driving pulse for the waves and oscillations in the ambient magnetic field, whose frequencies do not correspond to those of the driving reconnection. Forward modelling  the gyrosynchrotron radio emission associated with this process reveals that the QPPs may have multiple components,  determined  both by  the oscillatory reconnection and  the oscillations of the ambient flux rope.   Although our work is primarily motivated by solar flares, the simulations are quite generic and the results also have potential relevance to the merger of magnetic flux ropes in laboratory and space plasmas.

\section{Methodology}

\subsection{Numerical simulations}
We solve the two-dimensional resistive magnetohydrodynamic (MHD) equations using the LARE2D code, \new{version 4.2} \citep{Arber2001}. The equations, solved in the Lagrangian form using a Lagrangian-Eulerian remap procedure \citep{Arber2001},  are expressed in dimensionless form as follows:
\begin{eqnarray}
\frac{D\rho}{D t} &=& - \rho \nabla \cdot \mathbf{v} , \\
\rho\frac{D\mathbf{v}}{D t} &=& (\nabla \times \mathbf{B} ) \times \mathbf{B} - \nabla p + \mathbf{f}_{visc} , \\
\frac{D\mathbf{B}}{D t} &=& (\mathbf{B}\cdot\nabla) \mathbf{v} - \mathbf{B}(\nabla \cdot \mathbf{v}) - \nabla\times (\eta\nabla\times \mathbf{B}) , \\
\frac{D\epsilon}{D t} &=& - \frac{p}{\rho} \nabla \cdot \mathbf{v} + \frac{\eta}{\rho}\mathbf{j}^2 , \\
p &=& \rho \epsilon(\gamma -1) .
\label{eq:mhd}
\end{eqnarray}
Here,  $\mathbf{v}$ stands for the velocity vector, $\mathbf{B}$  the magnetic field, $\mathbf{j}$  the current density, $\rho$  plasma density, $p$ thermal pressure, $\epsilon$  specific internal energy, $\eta$  the resistivity, and $\gamma$ is the ratio of specific heats, which is set to 5/3 for hydrogen plasma. A numerical viscosity vector, $\mathbf{f}_{visc}$, is incorporated to address numerical instabilities and accommodate steep gradients such as shocks \cite{Arber2001,Bareford2015}.
The model considers the plasma to be  fully ionised, and the governing equations are normalised with respect to a length-scale, $L_0$, magnetic field, $B_0$ and density, $\rho_0$. These three fundamental normalising constants are then used to define the normalisation for velocity, $v_0 = B_0/\sqrt{\left.\mu_0 \rho\right.}$, pressure, $P_0=B_0^2/\mu_0$, time, $t_0=L_0/v_0$, current density, $j_0=B_0/(\mu_0 \rho)$, specific internal energy, $\epsilon_0=v_0^2$, and resistivity, $\eta_0= \mu_0L_0v_0$, scales. The simulation results can be scaled  with any appropriate reference scales, \new{which is done in Section IIC below where gyrosynchrotron radiation is considered}. The resistivity model is defined as a combination of a background resistivity, denoted as $\eta_b$, and a current-dependent anomalous resistivity, $\eta_0$, defined as
\begin{equation}
  \eta = 
\begin{cases}
  \eta_b + \eta_0,& |\mathbf{j}|\geq j_{crit} ,\\
  \eta_b,       & |\mathbf{j}|< j_{crit} ,
\end{cases}
\end{equation}
where $j_{crit}$ is a critical current adopted as 1.2 for every simulation (chosen so that only background resistivity is present in the initial conditions, but anomalous resistivity is switched on within current sheets). Anomalous resistivity is motivated physically by the likelihood of kinetic plasma instabilities occurring in regions of strong current, and is also commonly included in models of reconnection in the solar atmosphere in order to allow reconnection whilst minimising the effects.  global Ohmic dissipation.   Here, we consider the effects of varying both a constant resistivity, obtained by setting $\eta_0=0$ and varying $\eta_b$, and varying anomalous resistivity $\eta_0$. 

\subsection{The initial configuration}
\begin{figure}
\centering
\includegraphics[width=0.8\columnwidth]{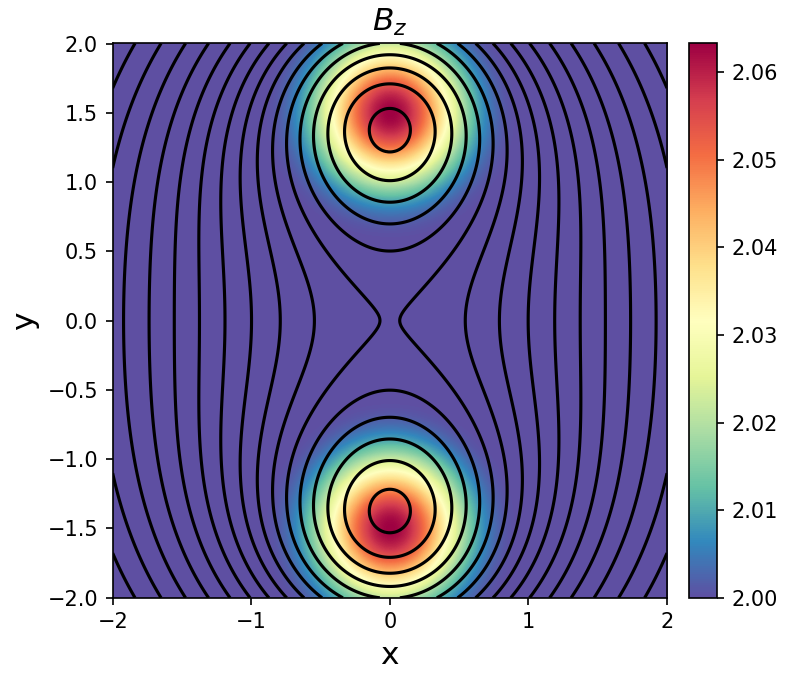}
\caption{Simulation initial conditions. The coloured contours show the magnetic field in $z$ direction, $B_z$, and the black lines represent the in-plane magnetic field lines.}
\label{fig:sethl}
\end{figure}

We consider a model of reconnection during the merger of two identical current-carrying twisted flux ropes, following \citet{Stewart2022} and based on earlier simulations of merging flux ropes in spherical tokamaks \cite{Stanier2013,Browning2016}. The initial configuration is 2D and  invariant in the z-direction, so that the flux ropes are straight cylinders. The initial configuration, described more fully in \citet{Stewart2022}, comprises two circular  flux ropes of radius $w$ centred at $z= \pm h/2$, with uniform axial field $B_{z0}$ outside the ropes. The initial condition of the simulation is illustrated in Fig. \ref{fig:sethl}, depicting field lines of the in-plane magnetic field as well as the  distribution of the axial magnetic field  $B_z$ (the "guide field"). The magnetic field lines are obtained solving a Poisson's equation for the stream function, $\psi$, as $\nabla^2\psi=-\mu_0 j_z$. Within each flux rope, the fields are defined to be a local force-free equilibrium but there is an attractive force between the flux ropes due to their carrying like currents (mathematically, this arises as the superposition of two force-free fields is not force-free).

Considering a cylindrical coordinate system where $(r,\phi,z)$ are the radial, polar, and axial directions, the initial current density for a single flux rope is described by
\begin{equation}
j_z = 
\begin{cases}
j_m \left(1 - \frac{r^2}{w^2} \right)^2 , & r\le w ,\\
0,& r > w \ ,
\end{cases}
\label{eq:current_density}
\end{equation}
\noindent where the peak of current is given by \new{$j_m=1$} at the flux rope centre, which decays to zero at the edges. The azimuthal component of the magnetic field, $B_{\phi}$, is obtained by applying the Amp\`{e}re's Law to Eq.\ (\ref{eq:current_density}), and it is given by
\begin{equation}
B_{\phi} =
\begin{cases}
B_p\left( \frac{3r}{w} - \frac{3r^3}{w^3}+\frac{r^5}{w^5}\right) , & r\le w ,\\
B_p \frac{w}{r}, & r > w ,
\end{cases}
\label{eq:inital_bp}
\end{equation}
\noindent  where $B_p$ quantifies the magnitude of the poloidal (in-plane) component of the magnetic field, given by $B_p = \frac{wj_m\mu_0}{6}$, where $\mu_0$ is the vacuum magnetic permeability. The axial magnetic field $B_z$ inside the flux rope is computed to make the Lorentz force zero for each individual flux rope, $\mathbf{j} \times \mathbf{B}= 0$. The axial magnetic field is defined by the following expression;
\begin{equation}
B_z = 
\begin{cases}
B_{z0} \left(1 +\frac{B_p^2}{B_{z0}^2} \mathcal{P}\right)^{1/2} , & r\le w ,\\
B_{z0} ,& r > w \ ,
\end{cases}
\label{eq:inital_bz}
\end{equation}
\begin{equation}
\mathcal{P} = 
\frac{47}{10} -
\frac{18r^2}{w^2} +
\frac{27r^4}{w^4} -
\frac{20r^6}{w^6} +
\frac{15r^8}{2w^8} -
\frac{6r^{10}}{5w^{10}} .
\end{equation}

The dimensionless quantity  $B_p/B_{z0}$ defines the ratio between poloidal and axial  magnetic fields; for a coronal loop, this ratio is typically  $\le 1$. The magnetic field defined by Eqs.\ (\ref{eq:inital_bp}),(\ref{eq:inital_bz}) for each flux rope is superposed generating the initial condition shown on Fig.\ \ref{fig:sethl}. Note that this configuration is not force-free \cite{Stanier2013,Stewart2022}, due to the non-linearity of the Lorentz force, and there is an attractive force between the two flux ropes (as "like currents attract") which causes the flux ropes to move together and eventually reconnect. 
In our simulations, the initial flux rope separation is set to $h=1$, and their radii as $w=1$. The initial density is uniform $\rho=1$ and the specific internal energy, $\epsilon=$0.01, corresponding to a low plasma $\beta$ as in the solar corona, and velocity is set  to zero. We do not consider here variations in $B_p/B_{z0}$, as this was investigated previously \cite{Stewart2022} and thus set  $B_{0z}$= 1 and \new{$B_{p}=1/6$}. 
The initial values of resistivity are detailed in table \ref{tab:simulation-setup}.

\begin{figure}
\centering
\includegraphics[width=0.8\columnwidth]{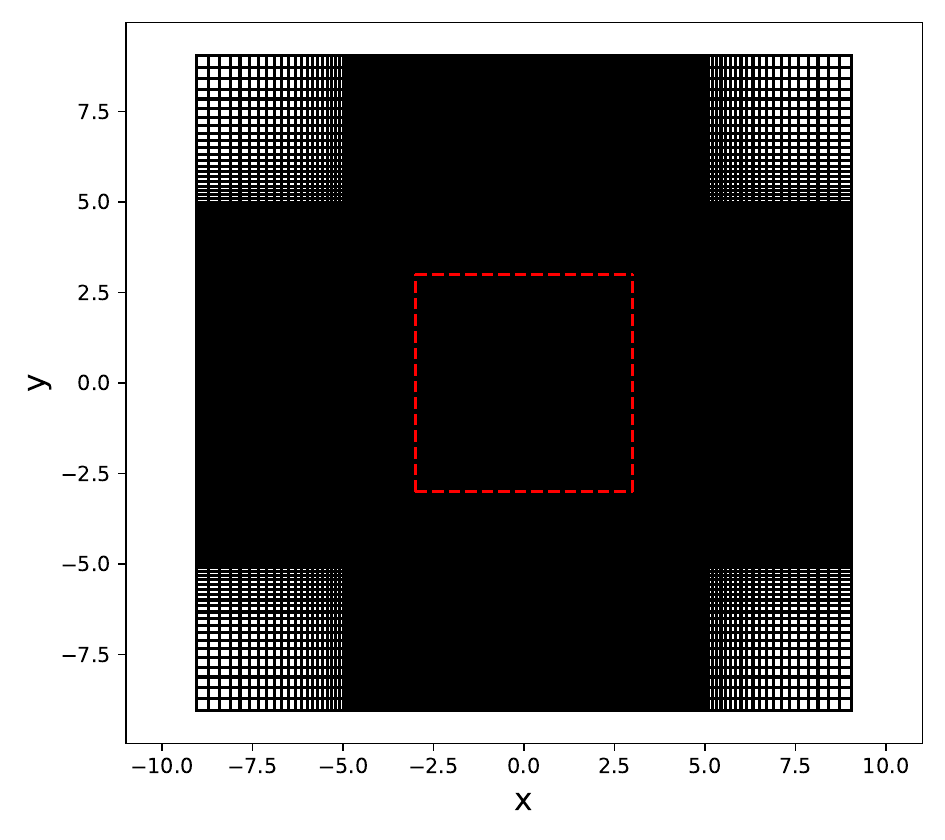}
\caption{The mesh configuration for grid 2. The red box indicates the region where the grid is equally spaced.}
\label{fig:grid}
\end{figure}

The original simulation model from \citet{Stewart2022} was updated in order to take advantage of some numerical improvements  available in LARE2D. First, we adopt far field boundary conditions, thus the variables at the boundaries are determined by the Riemann invariants, thus \new{minimising} wave reflections from the (unphysical) outer boundaries of the system.
We also employed a stretched grid characterised by finer resolution closer to the flux ropes, and coarser resolution in the outer regions, allowing an effective increase in resolution without need to increase the number of grid points. Figure \ref{fig:grid} illustrates the model of staggered grid used in our study. The region indicated by the red square is equally spaced and highly refined, between -3 and 3, placed in the part of domain where the magnetic reconnection takes place. For the external grid, we adopt a hyperbolic stretching function that smoothly changes the growth rate of grid spacing up to 7\% at the boundaries. The mesh stretching in the outer regions also creates some numerical dissipation, which is useful in terms of reducing the impact of reflected waves. 

\begin{table}
\centering
\caption{Definition of background and anomalous resistivities employed in each numerical simulation.}
\label{tab:simulation-setup}
\begin{tabular}{lcccccc}
\hline \hline
Case & $\eta_b$ & $\eta_0$  \\ \hline
A1 & 5$\times 10^{-4}$ & 0 \\
A2 & 1$\times 10^{-4}$ & 0 \\
A3 & 5$\times 10^{-5}$ & 0 \\
A4 & 1$\times 10^{-5}$ & 0 \\
B1 & 1$\times 10^{-5}$ & 1$\times 10^{-2}$ \\
B2 & 1$\times 10^{-5}$ & 5$\times 10^{-3}$ \\
B3 & 1$\times 10^{-5}$ & 1$\times 10^{-3}$ \\
B4 & 1$\times 10^{-5}$ & 5$\times 10^{-4}$ \\
\hline \hline
\end{tabular}
\end{table}

\subsection{Gyrosynchrotron analysis}
In order to determine the potential consequences for QPPs, we forward-model the microwave radiation from our system, with the intention of detecting any oscillatory behaviour. 
Mildly-relativistic electrons within a coronal loop gyrate in magnetic fields, emitting gyrosynchrotron (GS) radiation typically in the microwave frequency range during a solar flare. However, accurate calculation of GS radiation is computationally expensive. To address this, we use a fast GS radiative transfer code developed by \citet{Fleishman2010}\cite{Nita2015, Kuznetsov_2021}.

The radiative transfer code enables the user to take the number density (cm$^{-3}$), temperature (K), and magnetic field along a line-of-sight (T), to calculate the GS radiation intensity (in solar flux units) emitted along that line-of-sight for a range of selected frequencies. It significantly reduces the computational time required for calculating GS radiation, providing results within 1-10\% of their exact solutions. This algorithm has been previously used in the study of solar flares \cite{Chen2020, Kontar2017, Gordovskyy2017}, for investigating QPPs \cite{Smith2022, Altyntsev2016, Shi2023}, and has applications outside solar physics \cite{Climent2022, Waterfall2018}. Another noteworthy feature of the algorithm is its adaptability, allowing users to choose from a variety of electron energy and pitch-angle distributions or define their own. In this paper, we use a thermal energy distribution and an isotropic pitch-angle distribution.

We use the radiative transfer code to study and compare contributions to the GS emission from the reconnection site and the wave-propagating region, and to determine the effect of resistivity on these contributions. The code calculates these contributions in scaled units. We assign scaling factors for two merging plasmoids in a fragmented current sheet, following the values used in \citet{Stewart2022}: $B_0 = 0.0075$ T, $\rho_0 = 9\cdot 10^{-12}$ kgm$^{-3}$, $L_0 = 10$km. This gives a time scaling factor of $t_0 = 4.48\cdot 10^{-3}$s.  Since the time between each simulation snapshot is 1 Alfvén time, it is also $4.48\cdot 10^{-3}$s. At $t = 0$ these scaling factors give the following values for each plasmoid: a magnetic field strength of $|B| \sim 150$ G, in a background plasma with a density of $\rho \sim 10^{10}$ particles per cubic cm, and a background temperature of $T \sim 2\cdot 10^6$ K.

\section{Results}

\subsection{Grid convergence and validation}
\begin{figure}[h!]
\centering
\includegraphics[width=0.40\textwidth]{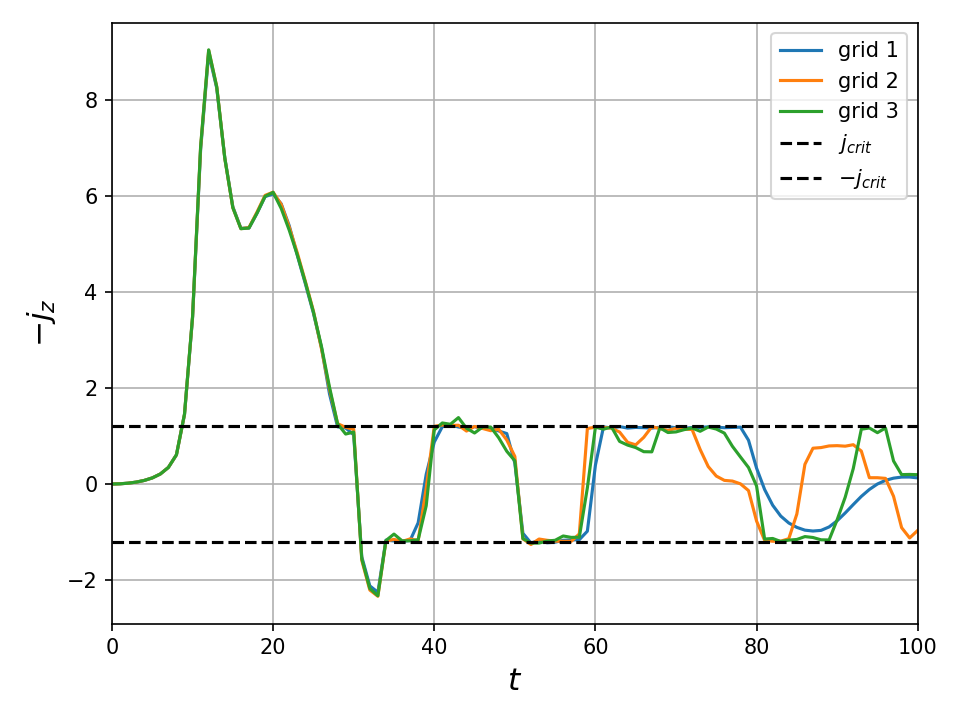}
\includegraphics[width=0.87\columnwidth]{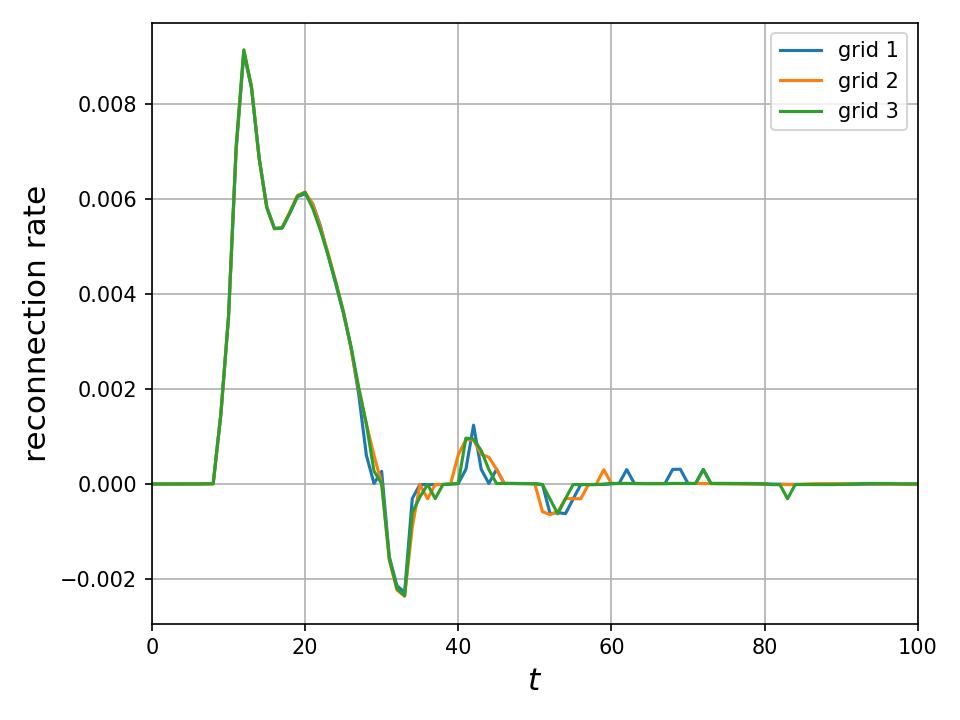}
\caption{The out-of-plane  current $j_z$ (a) and the reconnection rate (b) at the centre of the domain (0,0) as a function of time, for the simulation case B3 with different grid refinements (see Table II) }
\label{fig:grid-jz}
\end{figure}

To ascertain our results' reliability and grid independence, we conducted simulations for our baseline case, denoted as simulation setup B3 and detailed in Table \ref{tab:simulation-setup}, utilising three distinct levels of grid refinement. Figure \ref{fig:grid-jz} depicts two key aspects: (a) the vertical component of the current density at the X point, situated at the domain centre, and (b) the temporal evolution of the reconnection rate computed at the domain centre. These data are presented for three distinct grid refinement levels. Table \ref{tab:grid-setup} provides a detailed description of the grid configurations. The reconnection rate, discussed in depth by Comisso et al (2016) \cite{Comisso_2016}, is defined by the out-of-plane electric field, $-E_z$, measured at the X point at the origin \cite{Stanier2013,Stewart2022}. Given the symmetry of our system, the reconnection associated with the flux rope merger occurs primarily at the origin, although secondary reconnection may occur at other locations (see below). 
In the figures, the blue line corresponds to the grid resolution employed in \citet{Stewart2022}, while the orange and green lines represent two and four times more refined grids, respectively. Notably, all examined grid configurations exhibit consistent temporal behaviour, especially before $t=$ 48 (corresponding to 48 Alfvén times), and demonstrate very similar values. Some discrepancies in later time steps are acceptable, given the transient and highly non-linear nature of the problem.
The results collectively suggest that the grid with the lowest resolution can be judiciously employed without sacrificing the retrieval of vital information concerning the reconnection rate. Consequently, based on the outcomes of this test, we opt to utilise grid 2 in our subsequent simulations, thereby reducing the computational cost without compromising the accuracy of our results.
\begin{table}[h]
\centering
\caption{Grid setup employed in the validation studies.}
\label{tab:grid-setup}
\begin{tabular}{lccc}\hline \hline
Setup  & Grid points & Domain size   & Grid spacing   \\
  &  &  & in unstretched region \\ \hline
Grid 1 & 560 x 560   & -10.5 to 10.5 & 0.015625                                \\
Grid 2 & 950 x 950   & -9 to -9      & 0.0078125                               \\
Grid 3 & 1750 x 1750 & -7.5 to 7.5   & 0.00390625            \\ \hline \hline                 
\end{tabular}
\end{table}

\subsection{Magnetic field evolution and reconnection dynamics}

\begin{figure*}
\centering
\includegraphics[width=0.90\textwidth]{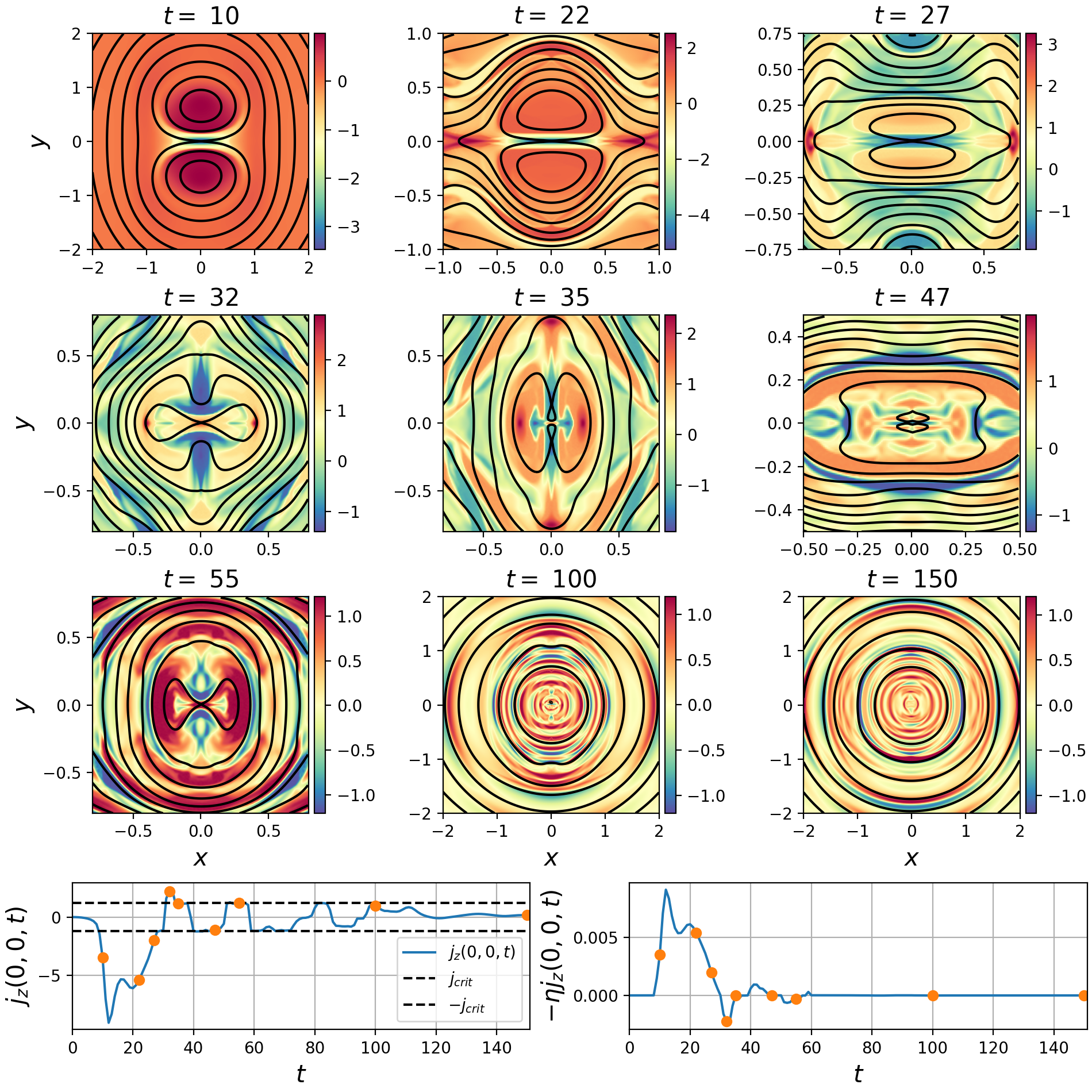}
\caption{Magnetic field evolution for simulation B3. The colour scale  illustrates the axial current density distribution, $j_z$, while the black lines represent the in-plane (poloidal) magnetic field lines. \new{The lower panels on the left-hand display the time variation of the current density, $j_z$, and the right the reconnection rate evaluated at the X-point. The red dots denote the time instants for the field plots above}.}
\label{fig:reconnection_evolution}
\end{figure*}

While our primary aim is to consider the effects of resistivity on the reconnection rate and oscillations (see Section C below),  and the properties of QPPs associated with the reconnection (see Section D), it is useful first to review the dynamics of the reconnection and oscillations.  The system's evolution can be divided into three main stages, each depicted in Figure \ref{fig:reconnection_evolution}.

The initial stage involves the collision and  merger of the flux ropes. During this phase, the flux ropes, initially separated, undergo acceleration and convergence due to the influence of the Lorentz force. Eventually, they merge to form a single flux rope \cite{Stanier2013}. 
The second phase initiates after the initial merger. In this stage, oscillatory reconnection phenomena manifest at the centre of the computational domain. This process generates a current density predominantly aligned with the x-axis, which periodically flips to an orientation along the y-axis \cite{Stewart2022}. These oscillations persist  but their  amplitude gradually diminishes. The dynamics is complex, with strong distortions of the in-plane field, and additional magnetic flux ropes being created and destroyed. 
The final stage follows the conclusion of the oscillatory reconnection process. During this phase, the single flux rope consists of approximately circular in-plane field lines which continue to oscillate, a behaviour initiated during the coalescence process in the first stage. Additionally, radially-outward propagating disturbances  continue to be emitted during this stage but with decaying  amplitude. Finally, the system approaches a new relaxed equilibrium, with lower magnetic energy than the initial double-flux-rope state, consisting of a single magnetic flux rope with circular cross-section \cite{Browning2014,Browning2016}. A more detailed description of the transient evolution can be found in \citet{Stewart2022}, and below. 

Figure \ref{fig:reconnection_evolution} illustrates  these stages of the transient evolution for case B3 (see Table I), showing the  out-of-plane current density, denoted as $j_z$ and  the in-plane magnetic field lines derived from isocountours of $\psi$, at successive time steps. The time-dependence of  current density, $j_z$, and the reconnection rate, $E_z$, at the origin are  also shown. At time $t = 10$, we observe the flux ropes approaching each other, resulting in an enhanced current density sheet, along $y=0$, evident  in the $j_z$ plot. Around time $t = 22$ the coalescence of the flux ropes is ongoing, \new{accompanied by the formation of jets along the $x$ axis. Subsequently, at t = 27, the islands are compressed, decreasing the reconnection rate before the current sheet's reorientation.}

Following this initial reconnection event, there is a clear  change in the orientation of the current sheet, which becomes evident around $t = 32$. At this point, there are two new flux ropes oriented in the $x$ direction. This arises from an "overshoot" of the reconnection process. At this stage the original flux ropes have merged completely (marking the end of the first phase), but there is still ongoing reconnection activity as we enter the second evolution phase. 
The oscillation of flux rope boundary compresses the two magnetic field islands observed in $t=32$ to a new configuration with two new X-points and extra small magnetic islands at $y=0$, as seen at $t=35$. By around $t=47$ the system returns to similar configuration observed at $t=27$ because of the compression of the magnetic field lines by the flux rope oscillation generated by the coalescence instability. Note the highly-distorted central magnetic island (associated with the reconnection outflow jets), and two small additional islands with centres aligned with the $y$ axis. 

The states at $t=$27, 32, and 35 represent a cycle that continues, seen in $t=47$ and 55, with pairs of islands forming and reconnecting with orientations successively parallel to $x$  and $y$ axes, but  eventually decays. The reorientation of the magnetic field lines is  correlated to the oscillatory pattern observed in $j_z$ and reconnection rate, $-\eta j_z$, displayed in the lower panels of Fig.\ \ref{fig:reconnection_evolution}.
The formation of magnetic islands visible for example at  $t = 22$ and $t = 35$ suggests that magnetic reconnection processes may occur both  at the origin  and in locations outside the central area. This implies complex magnetic reconnection which can be spatially distributed. To some extent, this is reminiscent of the familiar plasmoid instability, in which a current sheet breaks up into multiple islands \cite[e.g.]{Loureiro2007}, except that in our case the process is 2D, with islands forming with different orientations. 

From around  $t = $ 100  to $t = $ 150, the reconnection has finished, but ongoing oscillations can be observed. At this stage, the system has relaxed approximately to a single flux rope with circular in-plane field lines, which exhibit ongoing decaying oscillations and outward-propagating waves \cite{Stewart2022}. Similar magnetic field and current dynamics as displayed  in Fig.\ \ref{fig:reconnection_evolution} is also  observed for simulations B1, B2 and B4.

\begin{figure}
	\centering
	\includegraphics[width=0.98\columnwidth]{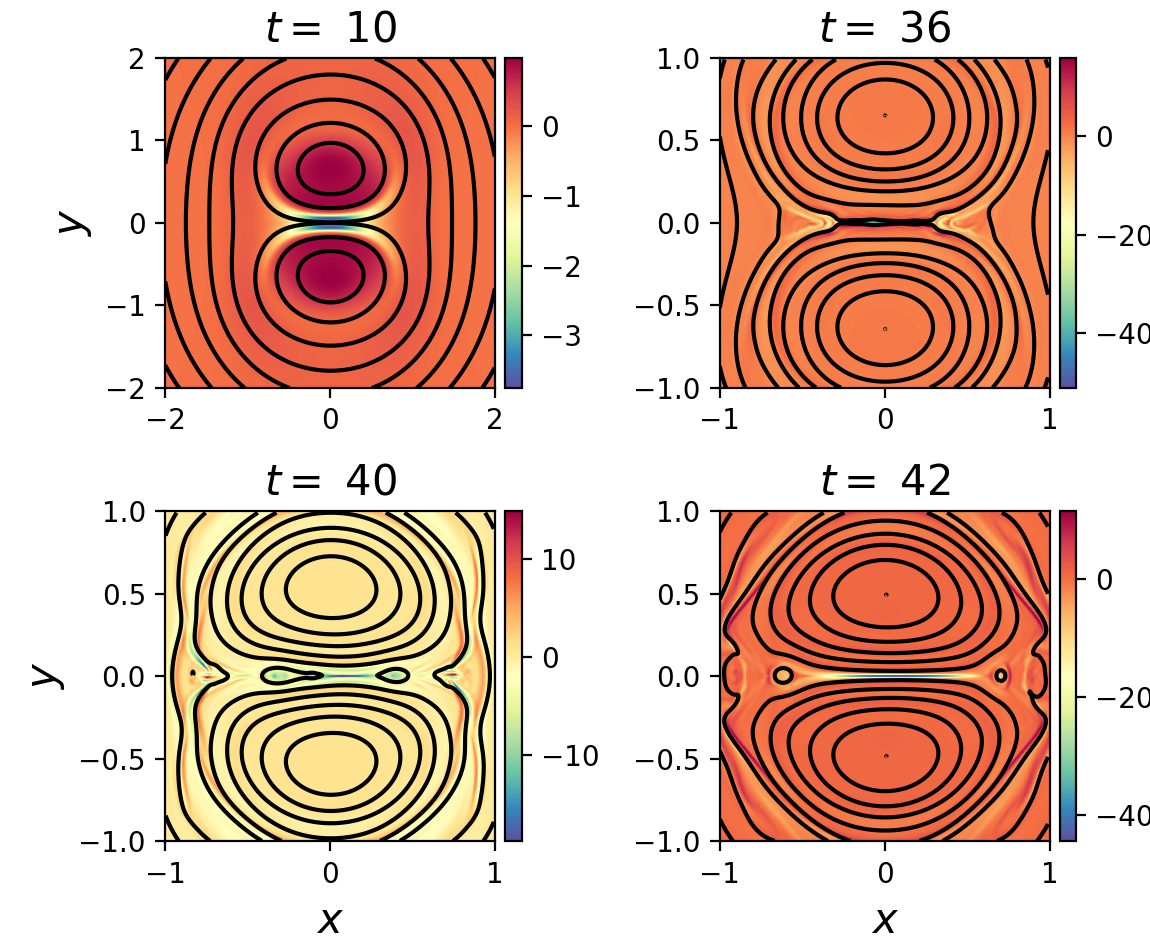}
	\caption{The magnetic field evolution for simulation A4 (constant resistivity case), see Table II. The colour scale shows the out-of-plane   current density distribution, $j_z$, while the black lines represent the magnetic field lines. }
	\label{fig:plasmoid}
\end{figure}

Simulation set A, where there is no anomalous resistivity, presents  different behaviour from simulation set B in some respects. Figure \ref{fig:plasmoid} shows the contour of $j_z$ and the magnetic field lines for simulation A4. This simulation set does not show the reorientation of magnetic field lines and current sheet presented in Fig.\ \ref{fig:reconnection_evolution}; instead of this, we observe  the formation of plasmoids in the current sheet  between the two initial magnetic field islands. The plasmoids start at $t=36$ with a small amplitude; they grow in $t=40$ and are ejected as seen in $t=42$. Interestingly, there is also symmetry breaking in the $x$ direction.

\begin{figure}
\centering
\includegraphics[width=0.8\columnwidth]{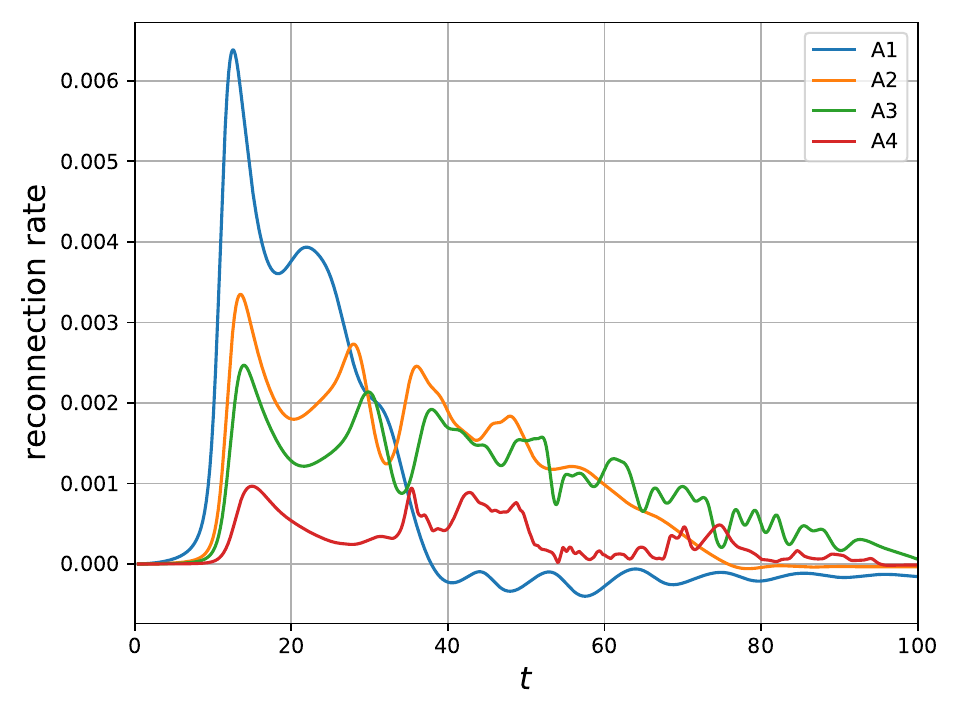}
\centering
\includegraphics[width=0.8\columnwidth]{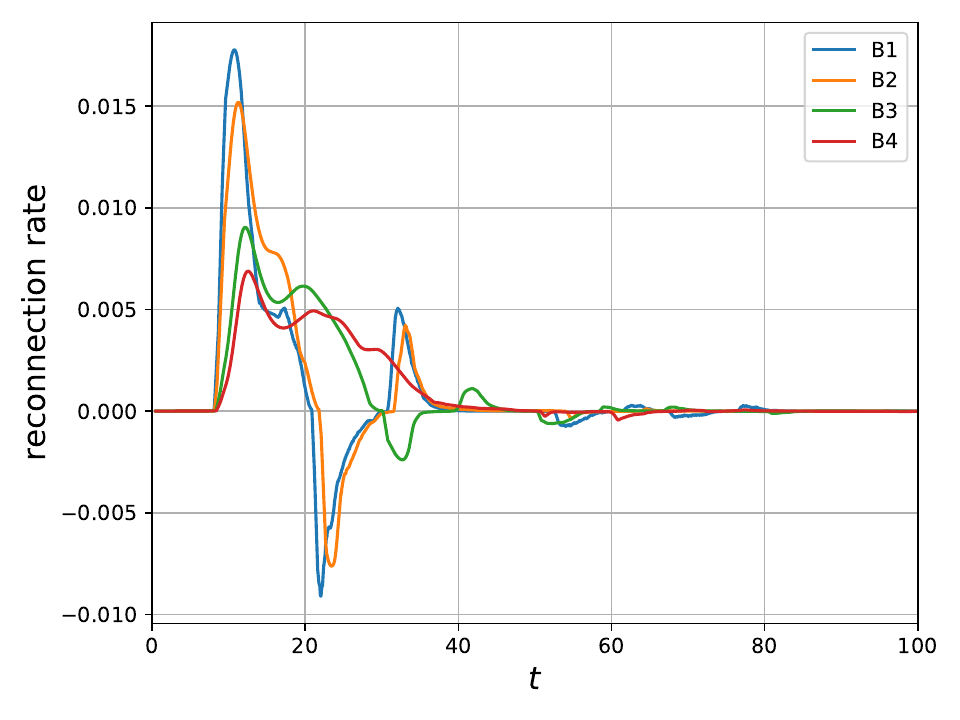}
\caption{Reconnection rate $E_z$ as function of time showing dependence on (a) uniform resistivity, $\eta_b$, for cases A1 to A4 (b) anomalous resistivity, $\eta_0$, for cases B1 to B4.}
\label{fig:rr-eta0}
\end{figure}

\begin{figure}
\centering
\includegraphics[width=0.99\columnwidth]{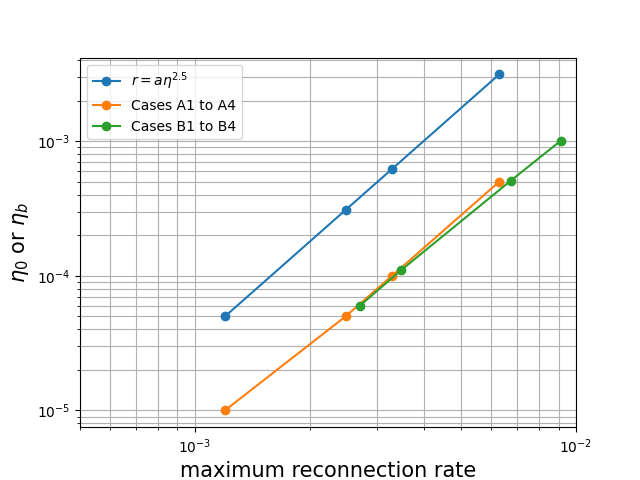}
\caption{Maximum reconnection rate as function of total resistivity, $\eta$, cases A1 to A4 and B1 to B4.}
\label{fig:loglog-eta}
\end{figure}

\begin{figure*}
\centering
\includegraphics[width=0.32\textwidth]{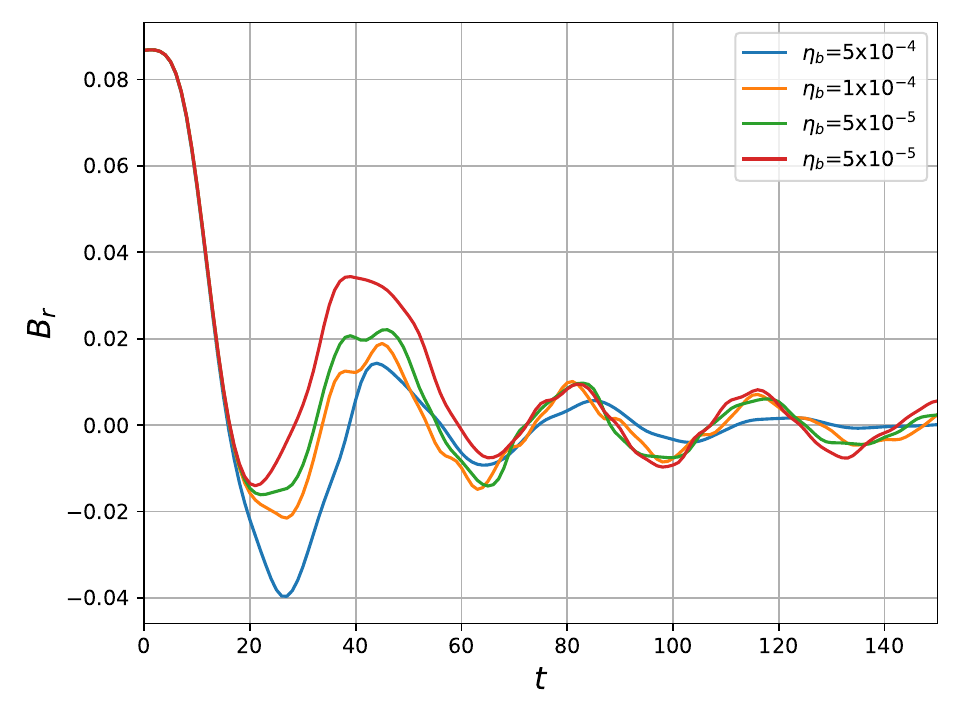}
\includegraphics[width=0.32\textwidth]{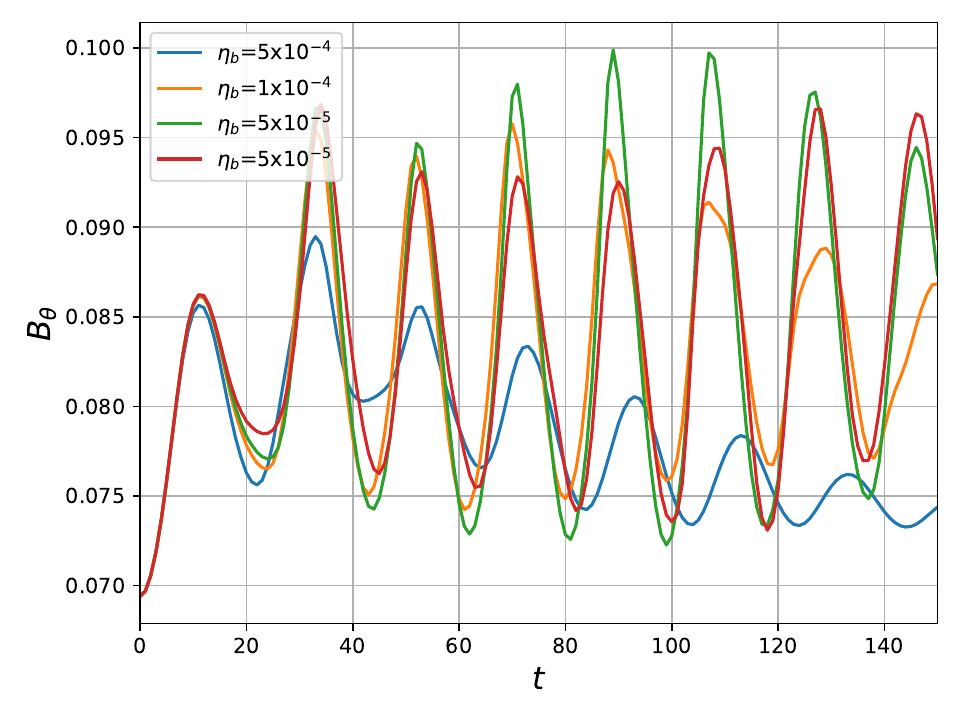}
\includegraphics[width=0.32\textwidth]{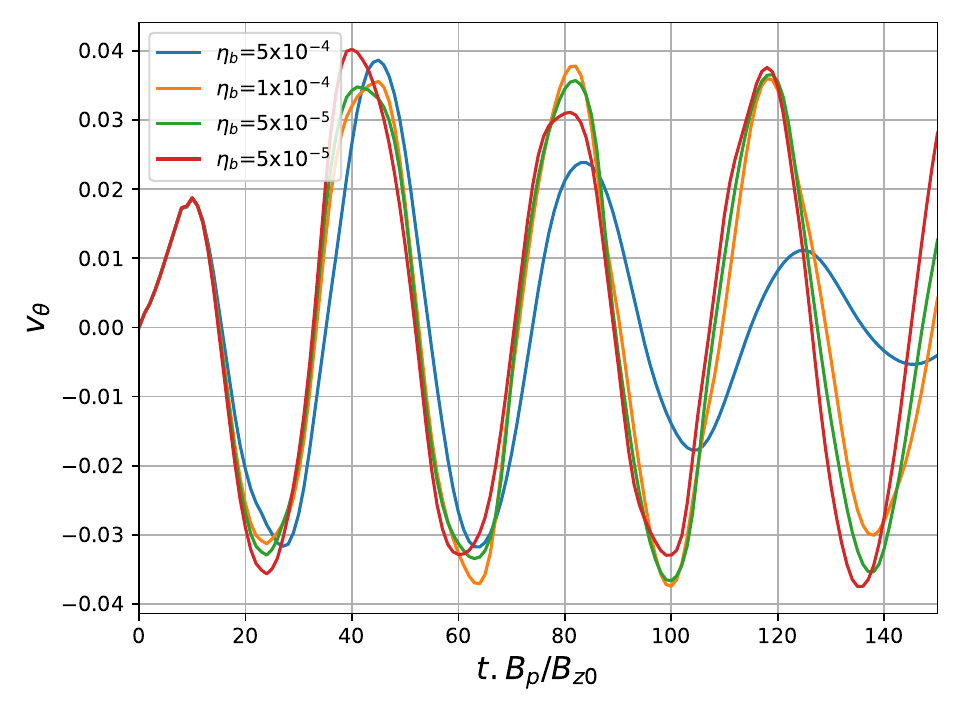}
\includegraphics[width=0.32\textwidth]{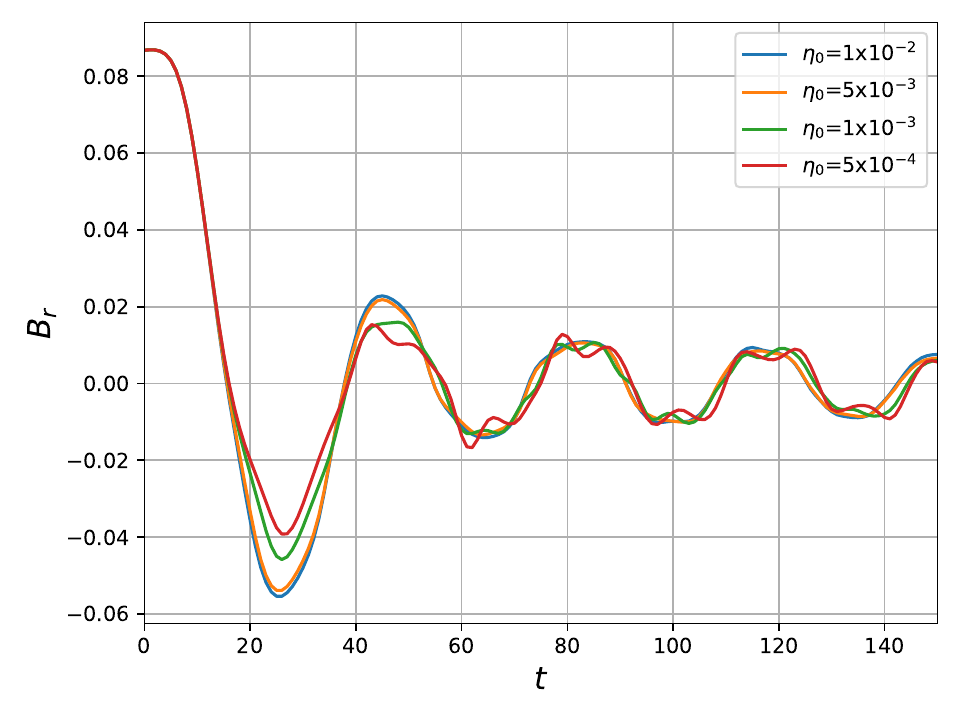}
\includegraphics[width=0.32\textwidth]{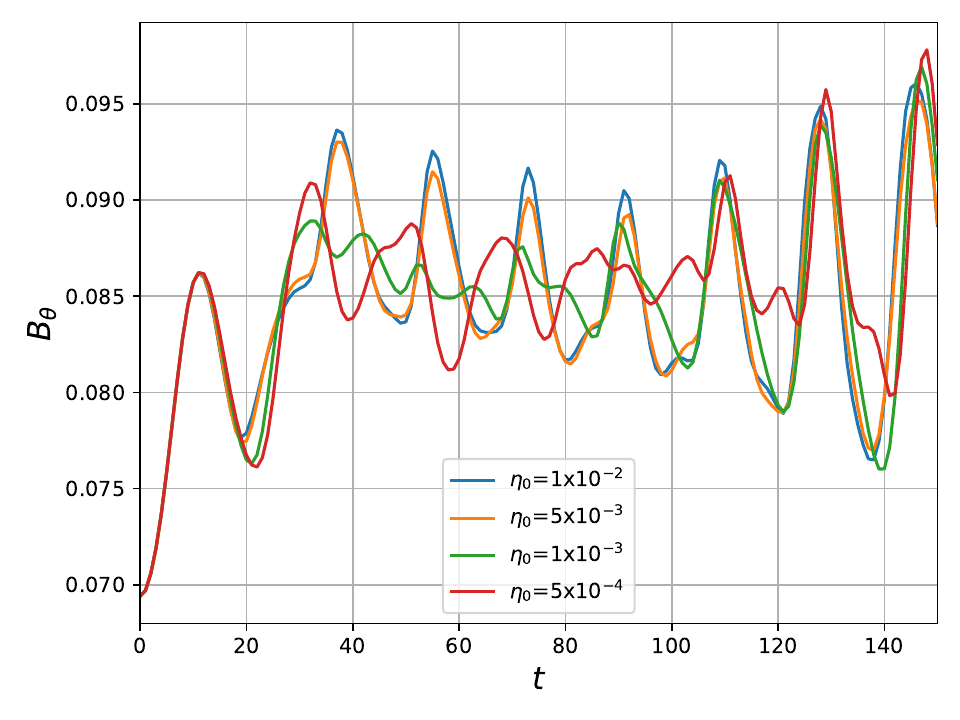}
\includegraphics[width=0.32\textwidth]{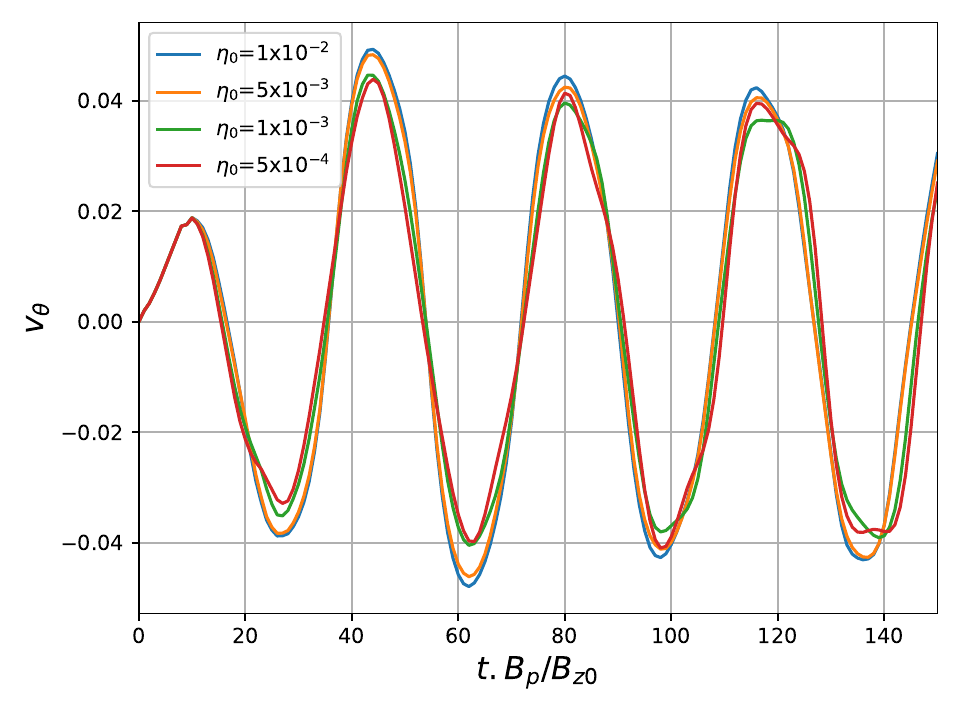}
\caption{The time evolution of $B_r$, $B_\theta$, and $v_\theta$, measured at $r=$ 1.8 and $\theta=$ 45$^o$. Upper line shows cases A1 to A4, and lower line B1 to B4 }
\label{fig:eta-variation}
\end{figure*}

\subsection{The effects of the resistivity profile}
We now explore the  influence of resistivity on the dynamics through a comparative analysis of two distinct scenarios:
(a) {\bf  Uniform resistivity} (background resistivity without anomalous resistivity), denoted as cases A1 to A4 in Table \ref{tab:simulation-setup}. 
(b) {\bf Anomalous resistivity} in conjunction with a small background resistivity, with different values of anomalous resistivity, represented as cases B1 to B4 in Table \ref{tab:simulation-setup}.

Figure \ref{fig:rr-eta0} illustrates the evolution of the reconnection rate for cases A1 to A4 and B1 to B4. A clear  trend emerges from examining cases A1 to A4 (uniform resistivity). As the  resistivity, $\eta_b$, increases, we observe, as expected  \cite{Stanier2013}, a  rise in the peak of the reconnection rate,  with the  most pronounced peak in the reconnection rate  observed at $\eta_b=5\times 10^{-4}$ (case A1). This case, with highest resistivity,   also exhibits a comparatively faster rate of decay in comparison to the other cases. As the resistivity decreases, the reconnection (although weaker) becomes more prolonged in time, and exhibits more oscillations. The lower resistivity cases (for example A4), also exhibit faster oscillations with multiple periodicities. These observations from simulations A1 to A4 suggest that both the peak magnitude of the reconnection rate, its subsequent decay and the character of the oscillations  are governed by the value of the resistivity, $\eta_b$. 

Similar trends are evident  when we analyse the effect of anomalous resistivity in cases B1 to B4. The peak of the reconnection rate increases as the value of the anomalous resistivity,  $\eta_0$, decreases. 
Figure \ref{fig:loglog-eta} illustrates the relationship between the maximum reconnection rate and the total resistivity, $\eta$. It is evident from the plot that the maximum reconnection rate  adheres to a power-law behaviour of the form $r=a\eta^b$. In this power-law relationship, the slope, represented by the exponent $b$, is approximately 2.5. It is worth emphasising that the scaling is almost identical (in terms of the total resistivity $\eta$) for both the constant resistivity and anomalous cases. This suggests that the reconnection rate is almost entirely determined by the local (potentially enhanced) resistivity within the current sheets.

For the anomalous resistivity cases, the subsequent time evolution of the reconnection rate also varies strongly with anomalous resistivity, with  reconnection becoming more prolonged, but with lower amplitude oscillations, as the resistivity decreases. The reversals in reconnection rate, noted by Stewart et al (2022) \cite{Stewart2022}, are most  evident for the cases with higher anomalous resistivity, B1 and B2, with a weaker reversal for case B3. Table \ref{tab:reconnection-integral} provides an overview of the time-integrated reconnection rate, over the duration of the simulation. It is notable that  the overall area enclosed by the reconnection rate curves, as depicted in Figure \ref{fig:rr-eta0}, which quantifies the net magnetic flux reconnected,  is very similar across the cases. This is consistent with the idea that the final state is a relaxed state or minimum energy state, more or less independent of the dynamical processes by which it is attained \cite{Browning2014,Browning2016}.
However, it is noteworthy that case A4, characterised by the smallest resistivity among all cases and lacking anomalous resistivity, stands out as an exception to this general trend. This may be due to the fact that reconnection activity is not fully completed during the time period of the simulations in this case.
\begin{table}[]
\centering
\caption{Result for the integration of reconnection rate module along the simulation time.}
\label{tab:reconnection-integral}
\begin{tabular}{lcccccccc}\hline \hline
Case     & A1    & A2    & A3    & A4    & B1    & B2    & B3    & B4    \\
$\int_0^T |j_z| \ \eta dt$ & 0.107 & 0.103 & 0.096 & 0.029 & 0.140 & 0.139 & 0.116 & 0.106 \\ \hline \hline
\end{tabular}
\end{table}

\begin{figure}
\centering
\includegraphics[width=0.99\columnwidth]{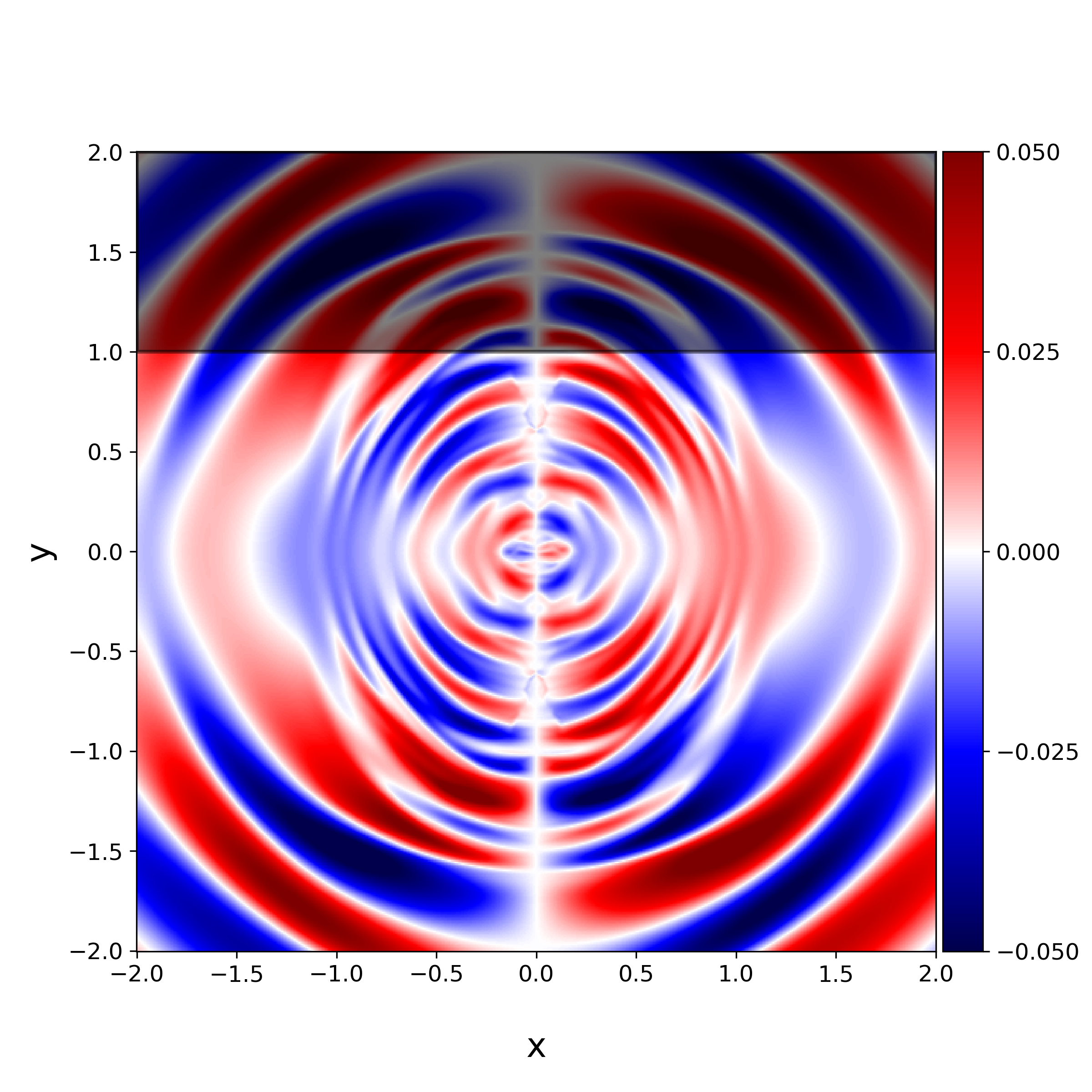}
\caption{Colour map of $v_x$ at $t = 100$ for simulation B3, illustrating the domain was divided for GS analysis. Domain A covers [-2:2, -2:2], while Domain B, represented by the shaded black box, covers [-2:2,1:2]}
\label{fig:domain-illustration}
\end{figure}

To analyse the impact of resistivity on emitted waves, we conducted a comparison using a "probe" located on the diagonal $x =y$ at $r=1.8$ (outside the reconnection region). Since the field geometry in the later stages of evolution is close to an  equilibrium with a single circular flux rope, it is convenient to consider the components of the magnetic field and velocity in cylindrical polar coordinates $r$ and $\theta$. Figure \ref{fig:eta-variation} presents measurements of $B_r$, $B_\theta$, and $v_\theta$ for two sets of simulations, cases A1 to A4
in the upper line, and cases B1 to B4 in the lower line. 
For cases A1 to A4, the plots corresponds to cases with variations in the background resistivity ($\eta_b$) from A1 to A4. 
For the cases with uniform resistivity, varying $\eta_b$, there is a damping effect on the emitted waves, associated with the fact that there is significant resistivity throughout the volume. Not surprisingly, damping is most pronounced in the case with the highest resistivity, A1. For cases A2 to A3, the damping is less significant. However, the frequency of the oscillations is almost constant  across these cases.

In the lower line (cases B1 to B4) figures corresponds to cases with variations in anomalous resistivity, $\eta_0$. Varying the anomalous resistivity  (B1 to B4) has minimal impact on the oscillations. Taken together with the cases A1-A4 above, there  observations suggest that variations in the background resistivity ($\eta_b$)  have a significant  effect on damping emitted waves, particularly at higher resistivity values. 
The anomalous  resistivity is only introduced  in cases B1-B4 when $|\mathbf{j}|<j_{crit}$, a condition primarily met near  the origin  and in a few neighbouring locations characterized by steep gradients in the magnetic field; hence, the waves in the outer regions (including the probe location) are only affected directly by the background resistivity.

However, the frequency of the emitted waves  - both for constant resistivity and with anomalous resistivity - is almost unaffected by the magnitude of the resistivity. This is an important result, as  - comparing with Figure \ref{fig:rr-eta0} - it shows that the time-dependence of the waves is unrelated to the time-dependence of the oscillatory reconnection. The emitted waves are likely to be natural oscillations of the final magnetic field configuration.





\begin{figure*}
\centering
\includegraphics[width=0.95\textwidth]{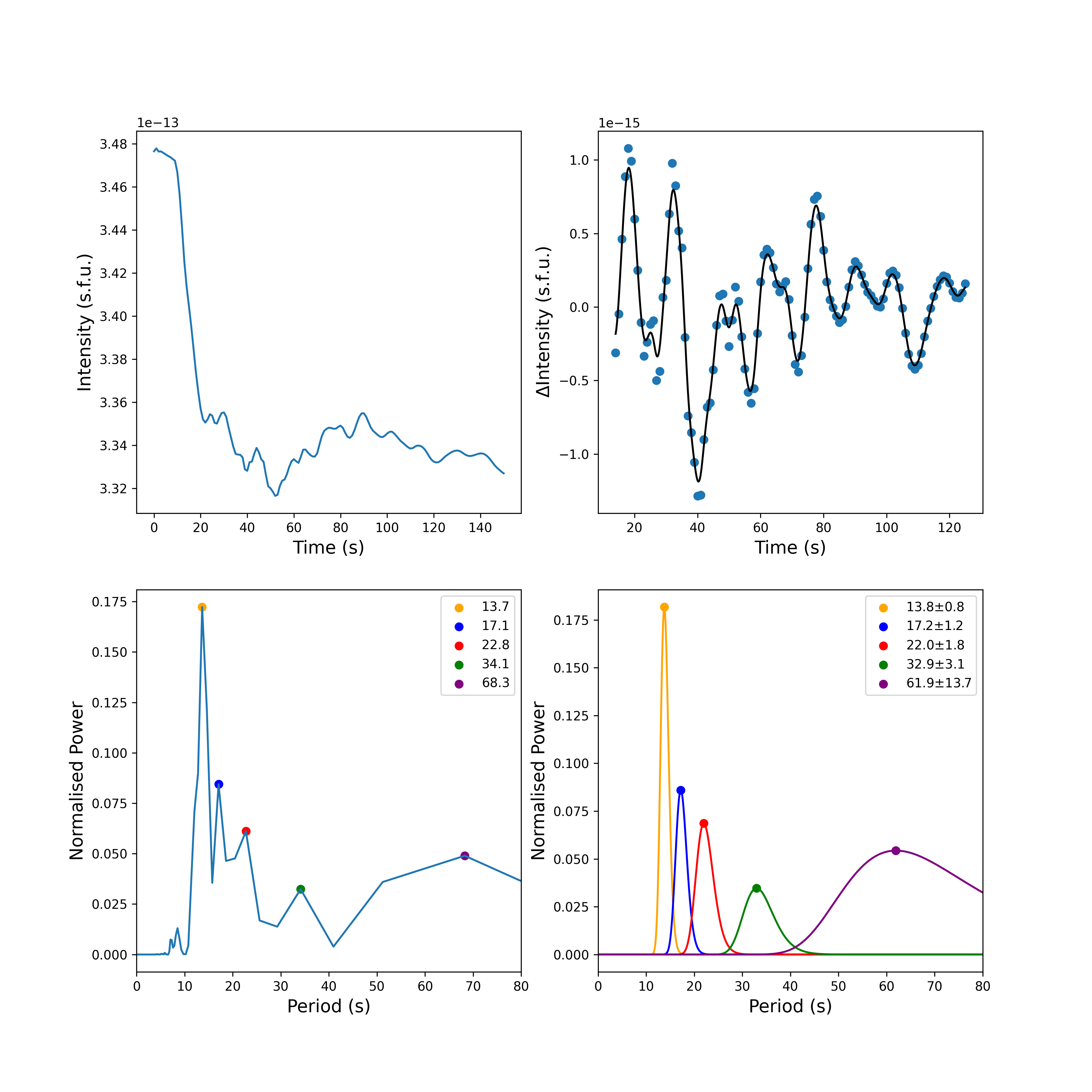}
\caption{\new{Analysis of oscillations in GS emissions for Domain A in Simulation B3. The top-left plot shows the intensity emitted from Domain A over time. The top-right plot presents the data with the moving average removed. The bottom-left plot displays the power spectrum of the emissions, while the bottom-right plot maps Gaussian peaks to the corresponding peak periods identified in the power spectrum.}}
\label{fig:intensity-B3-full-domain}
\end{figure*}

\begin{figure*}
\centering
\includegraphics[width=0.95\textwidth]{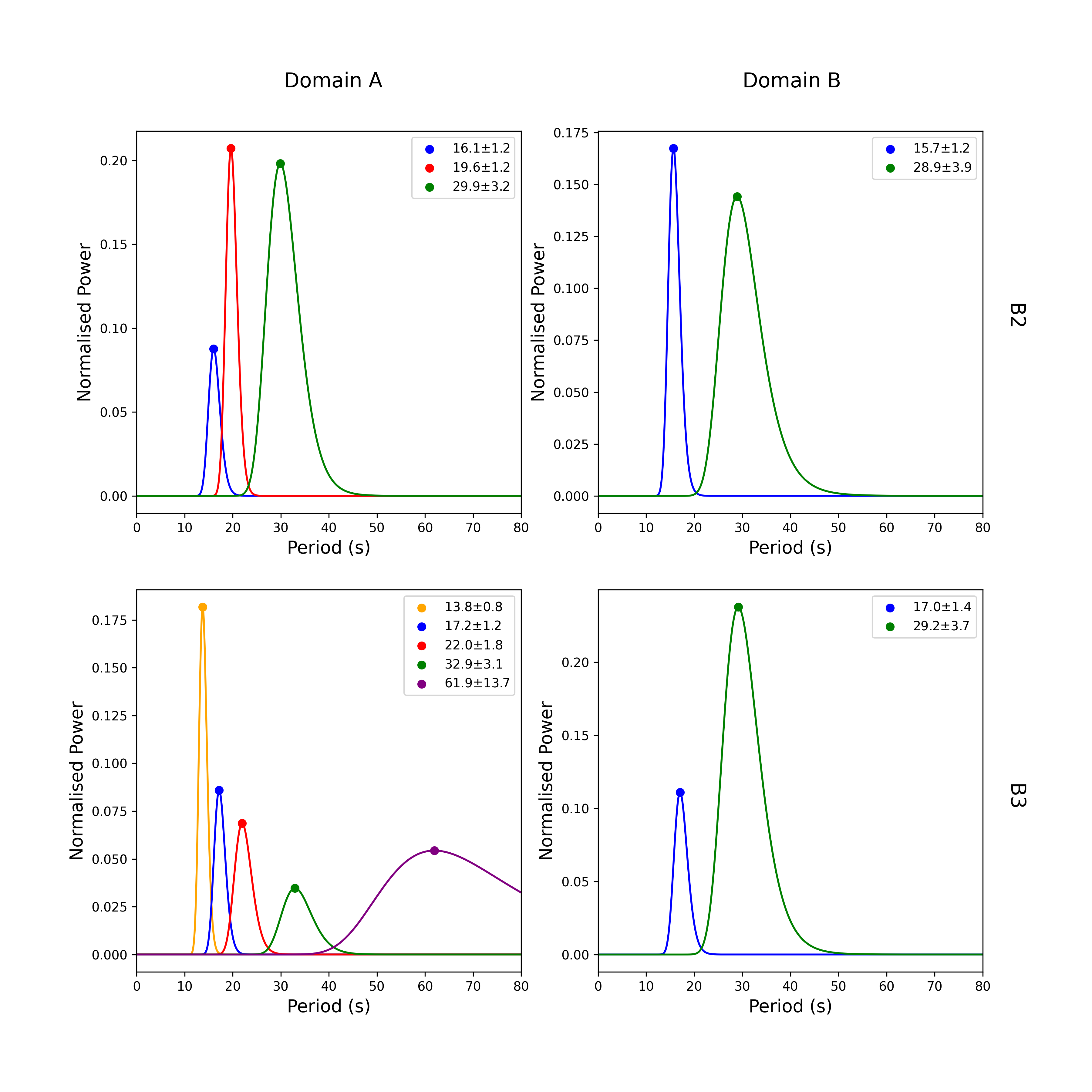}
\caption{\new{Comparison of peak periods in GS oscillations emitted from Domain A and Domain B for simulations B2 and B3. The first column shows peak periods in Domain A, while the second column shows peak periods in Domain B. The first row corresponds to simulation B2, and the second row to simulation B3. Peak periods found in wave propagating regions are coloured in blue and green, while those observed only at the reconnection site are coloured in yellow, red and purple.}}
\label{fig:intensity-comparison}
\end{figure*}

\subsection{Forward-modelling of the  emitted microwave radiation}
In order to assess the consequences of our results for the origin of QPPs in flares, we now forward-model oscillations in microwave (MW) radiation which would be observed from our model. 
To calculate the emitted radiation, we consider 508 lines-of-sight transverse to the loop in the x-direction. These lines are equally spaced within the region [-2,2:-2,2]. Along these lines, we record the density, temperature, and line-of-sight magnetic field. We then calculate the contribution to the GS emission from each line-of-sight. These individual intensities are then summed to obtain the total emitted intensity for each time step. This process is performed for both simulations B2 and B3, in order to determine the effects of varying the anomalous resistivity on the MW oscillations.

We use this technique to analyse two different domains, as illustrated in Figure \ref{fig:domain-illustration}. The first domain, Domain A, covers the entire [-2:2, -2:2] region, capturing both the reconnection site and the emitted wave region. Domain B is a sub-region [-2:2, 1:2], focusing on location of the emitted waves. Comparing emissions between Domain A and Domain B allows us to isolate the contributions to the GS emissions from the wave-propagating region and the reconnection site.

We detect, in both simulations, a peak in the GS emission around $2$ GHz. Oscillations are observed in the GS emission data. Figure \ref{fig:intensity-B3-full-domain} illustrates the evolution over time of the GS emission at $2$ GHz for Domain A in simulation B3, with similar behaviour observed at other MW frequencies. We investigate these oscillations further by removing a moving average from the original data and computing the power spectra of the oscillations. \new{The power spectra are then normalised so that the total integrated power equals 1. This allows for the comparison of the relative contributions of each peak within the same plot, though it should be noted that the magnitude of these peaks cannot be directly compared to those found in other plots. Using this method,} we detect several broad peaks for each oscillation, allowing us to identify their dominant period contributions. Calculating these periods may introduce spectral leakage and inaccuracies due to constraints in simulation length and temporal resolution. To address this, Gaussian peaks were fitted to each observed peak. The means of these curves provide a value for peak periods, while the variances serve as errors which allows for comparison with peaks from other power spectra.

The peak periods of GS emissions for Domains A and B in simulations B2 and B3 are depicted in Figure \ref{fig:intensity-comparison} and compared. In simulation B2, the peak periods from Domain A are approximately $16$, $20$, and $30$ seconds. The $\sim 16$ and $\sim 30$-second periods are also observed in the wave-propagating region. The period at around $20$ seconds is not, suggesting that it originates from an oscillation at the reconnection site. The total emitted radiation from the reconnecting flux ropes in simulation B2 contains contributions from both the reconnection site and the wave-propagation region.

In simulation B3 (with higher anomalous resistivity than B2),  two peaks are observed in Domain A: one at around $14$ seconds and another at around $17$ seconds, with the $\sim 14$-second peak being the dominant contribution. This indicates that resistivity influences the value of the peak periods present in GS emissions. In the wave-propagating region, peak periods are observed at around $17$ and $30$ seconds, but the $14$-second period is absent. This suggests that the $\sim 14$-second period originates from an oscillation at the reconnection site, similar to the B2 case. 

Notably, the oscillations emitted from Domain B are very similar between cases B2 and B3, showing that this component of MW emission is very insensitive to the resistivity. This is consistent with the fact that the magnetic and other disturbances  within this region are not greatly affected by resistivity (see above). There is however, still an important distinction between the B2 and B3 cases considering the entire domain. While the wave-propagating region contributes to the GS emissions across the entire domain in B3, its contribution is much lower than the $\sim 14$-second period originating from the reconnection site. Taking B2 into account, this suggests that when analysing GS emissions from two merging flux ropes, certain resistivities require the consideration of contributions from both the reconnection site and the emitted waves. However, in other cases, contributions from the wave-propagating region might not be significant.

\section{Discussion and conclusions}
We have carried out 2D resistive MHD simulations, based on a model developed by \citet{Stewart2022}, investigating  the merging of two current-carrying magnetic flux ropes through  oscillatory reconnection, and the generation of  slowly-propagating outward waves. For the first time,  forward-modelling of an oscillatory reconnection scenario has been undertaken in order to generate synthetic observables in microwave emission. Our results have important and novel implications for the origin of Quasi-Periodic Pulsations in solar flares.

Our first key finding is  that the outward-propagating waves emanating from the reconnection region are almost unaffected by the resistivity, and hence not directly correlated with the oscillatory reconnection.   The transient and oscillatory reconnection acts mainly as a driving pulse for oscillations in the surrounding field, which may be observed as QPPs. Our second key finding is a clear determination of  the relationship between observable QPPs and the reconnection and oscillations. The QPPs are (for our model), multi-component, with some periodicities determined by the magnetic reconnection and sensitive to the resistivity, with others arising from natural oscillations of the ambient magnetic field configuration and sensitive to the magnetic field and   plasma parameters. 

We have also considerably elucidated the dynamics of the reconnection, including its oscillatory nature  and its dependence on resistivity, and the  generation of waves in the surrounding field by the reconnection.
We explored the influence of resistivity by comparing two scenarios: one involving background resistivity without anomalous resistivity and the other incorporating anomalous resistivity alongside a small background resistivity. Our findings revealed that resistivity plays a critical role in controlling the rate of decay of the reconnection rate, directly affecting the oscillatory reconnection process. The maximum reconnection rate is demonstrated to be proportional to the total resistivity following a power-law. However, in contrast to the strong influence of resistivity on reconnection rates, we observed a different behaviour for the emitted waves. Background resistivity had a relatively minor damping effect on these waves. Importantly, the emitted wave frequencies remained almost constant across the cases we analysed.



We also investigated the GS emissions emitted from the simulations, isolating oscillations originating from the reconnection site and wave propagating region. By calculating the power spectra of these oscillations, we determined the peak periods contributing to each oscillation. Our findings indicate that both the reconnection site and wave propagating regions contribute to the oscillations in the GS emissions, with the contributions dependent on resistivity.

In the simulation B2, with lower anomalous resistivity, both regions contributed to the GS oscillations, whereas in the  more resistive case, B3, the reconnection site was the dominant contributor. This has implications for future simulations investigating the relationship between GS oscillations and the properties of merging flux ropes. It suggests that contributions from both the reconnection site and wave-propagating regions should be considered in some cases, while in others, contributions from the wave-propagating region may not be significant, depending on resistivity.

These findings highlight the complex interplay between resistivity, reconnection rates, and wave generation in the context of magnetic flux rope merging and oscillatory reconnection simulations. We have focused on merger of flux ropes which may occur in solar flares, either through coalescence of plasmoids  within a large-scale flare current sheet or through interactions of twisted coronal loops. However, flux rope merger is a common process in many space, astrophysical  and laboratory plasmas, and our results have broader implications: for example, to merging-compression formation in spherical tokamaks.

In a forthcoming paper we will present a more detailed analysis of the dynamics of the oscillatory reconnection and the nature of the waves generated by the reconnection,  including their dependence on the field and plasma parameters,  using modal decomposition tools. In future, it would also  be important to explore how the present results are modified in 3D configurations. 

\section*{Acknowledgements}

We wish to thank the UK Science and Technology Facilities Council (STFC) for providing studentship support for JS. PKB and LACAS are funded by the STFC grant ST/T00035X/1, LACAS also is funded by the STFC ST/X001008/1. We thank the Distributed Research utilising Advanced Computing (DiRAC) group for providing the computational facilities used to run the simulations in this paper.

\section*{Data Availability}

The data underlying this article will be shared on reasonable request
to the corresponding author (LACAS).
\bibliography{bibliography}


\end{document}